\documentclass{article}

\usepackage[final, nonatbib]{neurips_2024}

\usepackage{amsmath}

\usepackage[utf8]{inputenc} 
\usepackage[T1]{fontenc}    
\usepackage{hyperref}       
\usepackage{url}            
\usepackage{booktabs}       
\usepackage{amsfonts}       
\usepackage{nicefrac}       
\usepackage{microtype}      
\usepackage{xcolor}         
\usepackage{subcaption}
\usepackage{graphicx}
\usepackage{pgffor} 
\usepackage{float}  
\usepackage{xstring} 

\title{Are Grid Cells Hexagonal for Performance or by Convenience?}

\author{%
  Taahaa Mir, Peipei Yao, Kateri Duranceau\thanks{Code is available at: \texttt{https://github.com/tmir00/TemporalNeuroAI}} \\
  McGill University\\
  \texttt{\{taahaa.mir,peipei.yao,kateri.duranceau\}@mail.mcgill.ca} \\
  \And
  Isabeau Prémont-Schwarz\thanks{Corresponding author} \\
  McGill University\\
  \texttt{isabeau.premont-schwarz@mcgill.ca} \\
}

\begin{document}

\maketitle

\begin{abstract}
This paper investigates whether the hexagonal structure of grid cells provides any performance benefits or if it merely represents a biologically convenient configuration. Utilizing the Vector-HaSH content addressable memory model as a model of the grid cell -- place cell network of the mammalian brain, we compare the performance of square and hexagonal grid cells in tasks of storing and retrieving spatial memories. Our experiments across different path types, path lengths and grid configurations, reveal that hexagonal grid cells perform similarly to square grid cells with respect to spatial representation and memory recall. Our results show comparable accuracy and robustness across different datasets and noise levels on images to recall. These findings suggest that the brain's use of hexagonal grids may be more a matter of biological convenience and ease of implementation rather than because they provide superior performance over square grid cells (which are easier to implement in silico).
\end{abstract}

\section{Introduction}

The entorhinal cortex of the brain contains specialized neurons known as grid cells\cite{fiete2008what, mathis2012optimal}, which map our environment to assist with spatial navigation and memory recall. Biologically, these each of those grid cells fires in a regular pattern as the animal moves through space. That pattern forms a hexagonal grids. From a computational perspective, square-shaped grid tilings are often easier to implement. A key question that arises when trying to replicate these cognitive functions in silico is whether hexagonal grid cells provide better performance than square grid cells. If they do, we will want to put in the extra effort to implement them rather than the more conveniently implemented square grid cells. The biological prevalence of hexagonal grid cells \cite{hafting2005microstructure} suggests potential advantages of this particular tiling over other tilings, but if performance improvements are not evident, the reason for the hexagonal shape could be attributed to developmental simplicity rather than computational superiority.

To explore this hypothesis, we implement the Vector-HaSH \cite{Vector-HaSH} framework which is a neuroscientifically plausible computational model of the entorhinalcortex-hippocampal grid cells and place cell memory network (see section \ref{methods} for details about our implementation). We implemented two types of grid cells. The first tiles the plane with hexagonal tiles as is observed in the brain. The second tiles the plane with square tiles, as is much easier to implement in a computer. By comparing the performance of these structures in the context of storing and recalling memories, we aim to shed light on the relative advantages of each. Our evaluation uses the image datasets of MNIST, FashionMNIST and CIFAR-100, as memories to store and retrieve. Our simulated animal will also travel along three different types of paths: straight lines, Brownian motion, and Lévy flights, the latter being a better approximation to typical animal foraging behaviour than the first two.

This study not only aims to determine the effectiveness of hexagonal versus square grid cells but also contributes to a deeper understanding of how grid cell structures influence memory storage and recall in spatial tasks. The findings could have significant implications for the design of artificial cognitive systems.

In this paper, we investigate whether hexagonal grid cells offer inherent performance advantages or if they are simply a biologically convenient choice for spatial memory tasks. Our experiments, detailed in Section \ref{Experiments}, demonstrate that hexagonal grid cells perform similarly to square grid cells across a range of scenarios in image recognition and path simulations. This suggests that, at least for the tasks examined, the use of hexagonal grids may be more related to biological ease of implementation. The methodology for testing different grid configurations is outlined in Section \ref{methods}, and a thorough analysis of the results is provided in Section \ref{results_analysis}.

\section{Related Works}
Our work was significantly inspired by two experimental models based on the biological phenomenon of grid cells: the Memory Scaffold with Heteroassociation (MESH) model \cite{MESH} and the Vector Hippocampal Scaffolded Heteroassociative Memory (Vector-HaSH) model \cite{Vector-HaSH}. The Vector-HaSH builds on MESH by enhancing its original Content-addressable memory (CAM) network.

\subsection{MESH: Memory Scaffold with Heteroassociation}
This paper introduces a novel content-addressable memory (CAM) architecture called Memory Scaffold with Heteroassociation (MESH), designed to overcome the limitations of existing CAM networks, particularly the \textit{memory cliff}—a point at which adding a single additional pattern causes the catastrophic loss of all stored patterns. Inspired by the Entorhinal-Hippocampal memory system in the brain, MESH separates memory storage into two components: a \textit{memory scaffold} of predefined, stabilized states, and a \textit{heteroassociative} process that links these states to external patterns \cite{MESH}.
MESH achieves near-optimal information storage for any number of patterns, outperforming traditional CAM models \cite{MESH}. The paper also discusses the theoretical foundation of CAM networks, presents central results on MESH's performance, and extends the model to continuous neural activations.

\subsection{Vector-HaSH: Vector Hippocampal Scaffolded Heteroassociative Memory}
The Vector-HaSH model proposes a neocortical-entorhinal-hippocampal network that supports associative, spatial, and episodic memory \cite{Vector-HaSH}. It uses grid cells interacting with hippocampal cells via a scaffold of fixed and random projections, forming stable fixed points for high-capacity memory storage and recall. Key features include generalization from a small set of learned states, high-capacity storage without a memory cliff, and accurate recall even with partial inputs.

The model leverages interconnected layers—Input, Grid Cell, Hippocampal Cell, and a Scaffold—to maintain stable memory representations. Compared to MESH \cite{MESH}, Vector-HaSH provides superior generalization, resilience to interference, and increased sequence capacity. The grid cells are represented as a one-hot encoded vector on a discretized hexagonal lattice, allowing for more accurate memory encoding, improved pattern recall, and better sequence learning, making Vector-HaSH a more advanced model than MESH. This forms the basis of the memory model we use to assess the behaviour of Hexagon and Square grids in memory recall.

\section{Experiments}\label{Experiments}

\subsection{Image Datasets \& Evaluation}
To evaluate the performance of square versus hexagonal grid cells in the Vector-HaSH model, we conducted tests using three image datasets—MNIST, Fashion-MNIST, and CIFAR-100. When testing with a dataset, we sampled an equal amount of random images from each class depending on the number of total images $N$ to store, and stored them in our memory model at regular intervals along a virtual path through space traversed by the animal.

\begin{figure}[!htbp]
    \centering
    \includegraphics[width=0.9\linewidth]{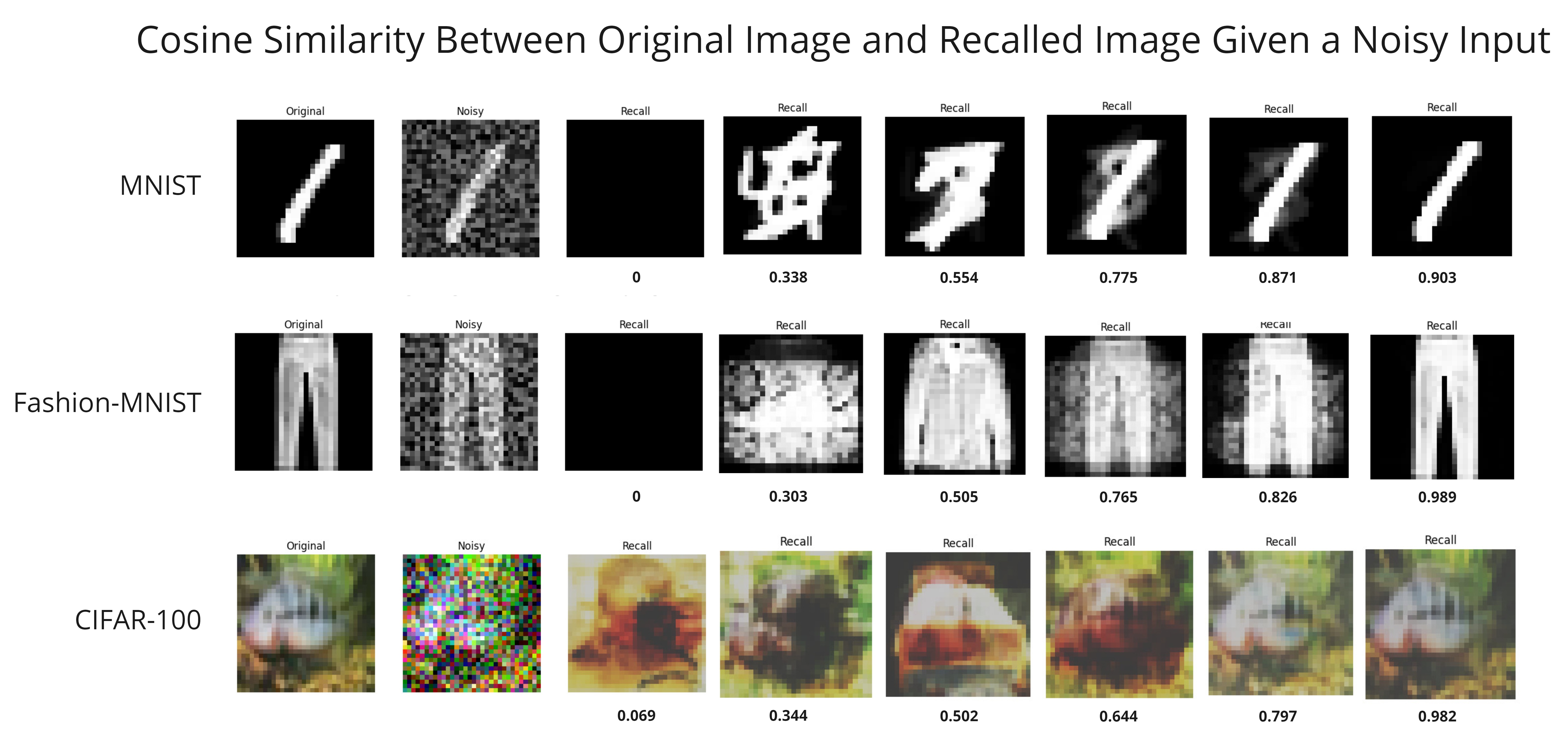}
    \caption{Examples of recalled images using a noisy input for all three datasets. The cosine similarity between the recalled image and the original is written below the recalled image.}
    \label{fig:cosinesim-examples}
\end{figure}

During recall, we introduced uniform noise to the pixels of all three datasets at varying intensities: None, low (ranging from -1 to 1), medium (-1.25 to 1.25), and high (-1.5 to 1.5). This added noise allowed us to test how effectively the models could reconstruct and recall the original images under different levels of distortion. The success of a recall was measured with the cosine similarity score between the recalled image and the original image (see Fig. \ref{fig:cosinesim-examples} for examples of recall quality as a function of cosine similarity for the different datasets).

Images were whitened before storing in the memory:
\[
x \leftarrow \frac{x - \text{mean}}{\text{std}},
\]

where \(x\) is an R (or G, or B) value of each pixel, \(\text{'mean'}\) is the computed mean over the dataset, and \(\text{'std'}\) is the standard deviation. During decoding, the transformation was reversed to recover the original pixel values in the RGB space for display.

We stored these images in our memory model and recalled them using noisy queries. The recall was assessed using the cosine similarity between the original and recalled images changes.

These comparisons with cosine similarity between original and recalled images provide a quantitative measure of accuracy, highlighting how effectively the model could denoise inputs and preserve the integrity of stored information.

\subsection{Path Simulations}
To determine with which active grid cells each memory should be stored, we simulated a path which the animal would take through space took points along that path at regular intervals. We then stored the memories with the grid cells which would have been active had the animal been at that physical location. We generated a series of coordinates corresponding to three distinct motion types to store memories: Straight Regular Line (Figure \ref{fig:straight_line_example}), Brownian Motion (Figure \ref{fig:brownian_motion_example}), Lévy Flight (Figure \ref{fig:levy_walk_example}):

\begin{figure}[h!]
    \centering

    \begin{subfigure}{0.3\textwidth}
        \centering
        \includegraphics[width=\textwidth]{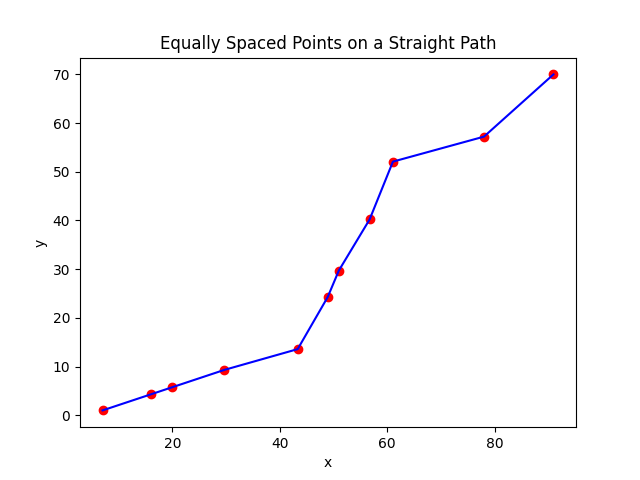}
        \caption{Represents a linear path.}
        \label{fig:straight_line_example}
    \end{subfigure}
    \hfill
    \begin{subfigure}{0.3\textwidth}
        \centering
        \includegraphics[width=\textwidth]{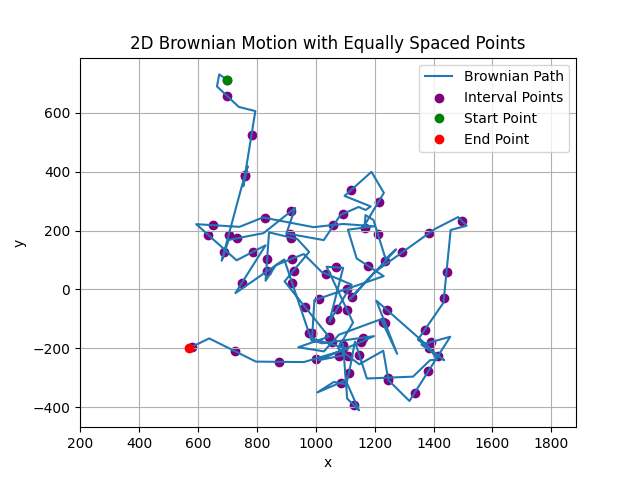}
        \caption{2D Brownian Motion.}
        \label{fig:brownian_motion_example}
    \end{subfigure}
    \hfill
    \begin{subfigure}{0.3\textwidth}
        \centering
        \includegraphics[width=\textwidth]{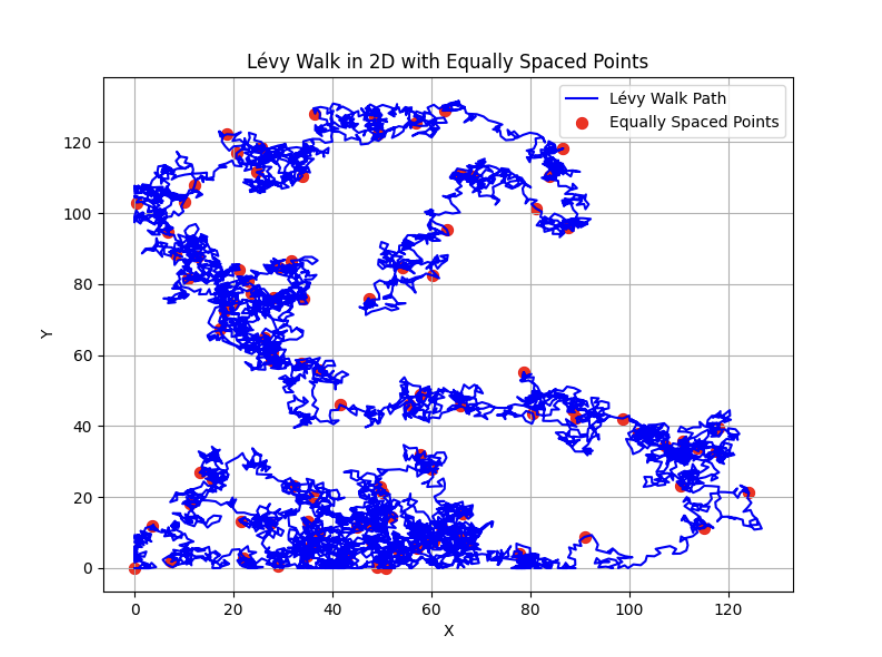}
        \caption{Lévy Flight Path.}
        \label{fig:levy_walk_example}
    \end{subfigure}
    
    \caption{Examples of the different path types used: regular line, Brownian motion, and Lévy flight.}
    \label{fig:path_comparison}
\end{figure}

These coordinates were mapped onto grids composed of either square or hexagonal cells, forming a grid vector $g$ that defines the storage locations for images from our datasets.

We also varied the length of the path across the three path types as follows:
\begin{verbatim}
    path_length = [N, N ** 2, N ** 3, N ** 4]
\end{verbatim}
where $N$ is the number of images stored in the Vector Hash model. As $path\_length$ increases and $N$ is unchanged, we have that each coordinate is further apart from the other. Note that all points are equally equally spaced along the trajectory.

\subsection{Results \& Discussion} \label{results_analysis}

In our experiments, we evaluated the performance of square and hexagonal grid cells across various image datasets and path simulations. The results reveal distinct performance trends depending on the number of stored images, noise levels, grid sizes, and path types, but all showed that Square Grid cells display similar performance to Hexagon Grid Cells.

 \textbf{Experiment 1: Varying amounts of stored images:} 

 \begin{figure}[!htbp]
    \centering
    
    \begin{subfigure}{0.45\textwidth}
        \centering
        \includegraphics[width=\textwidth]{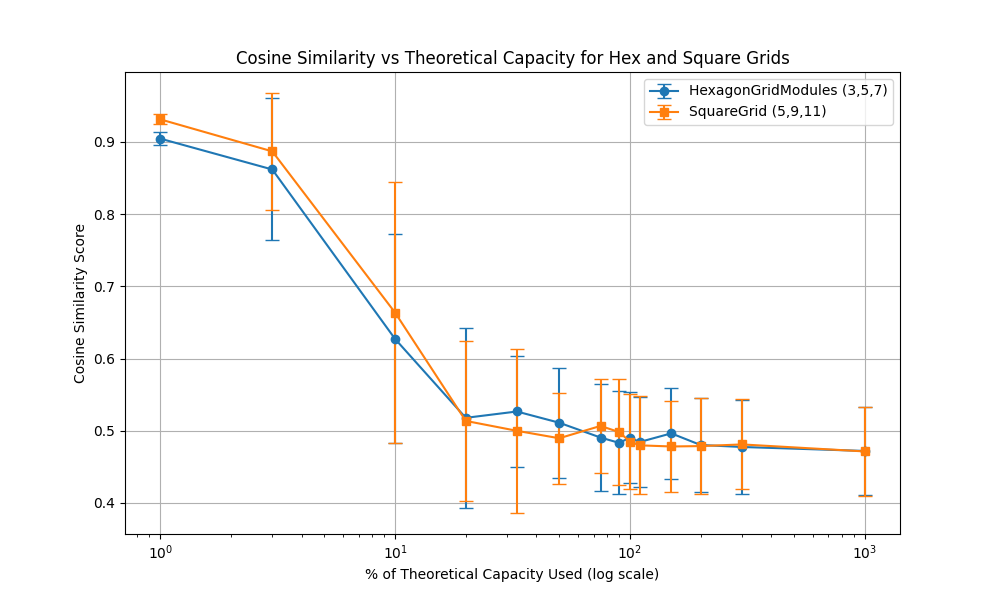}
        \caption{MNIST Images}
        \label{fig:NumImageMNIST}
    \end{subfigure}
    \begin{subfigure}{0.45\textwidth}
        \centering
        \includegraphics[width=\textwidth]{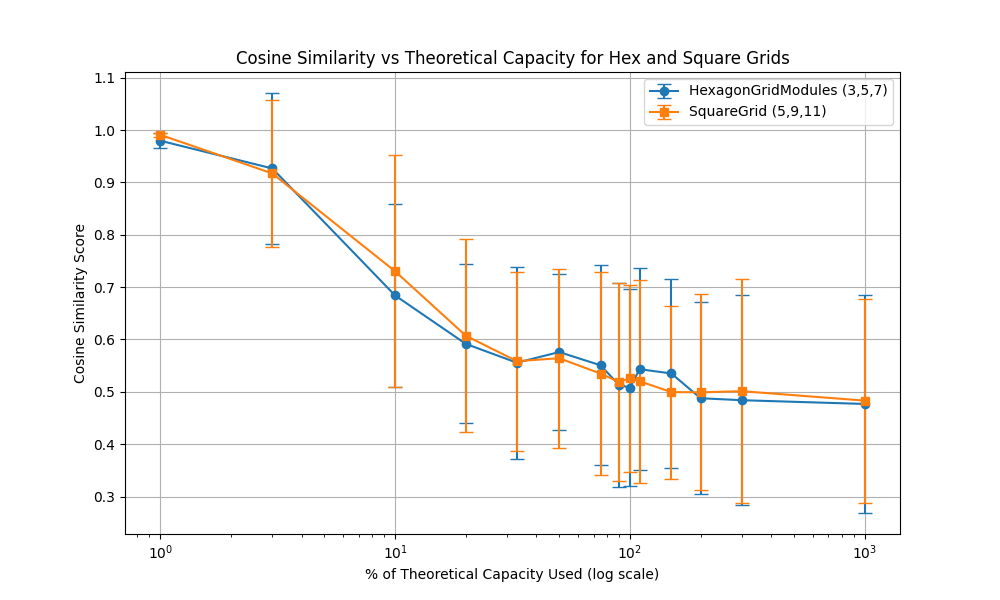}
        \caption{Fashion MNIST Images}
        \label{fig:NumImageFashionMNIST}
    \end{subfigure}
    \hfill
    \begin{subfigure}{0.5\textwidth}
        \centering
        \includegraphics[width=\textwidth]{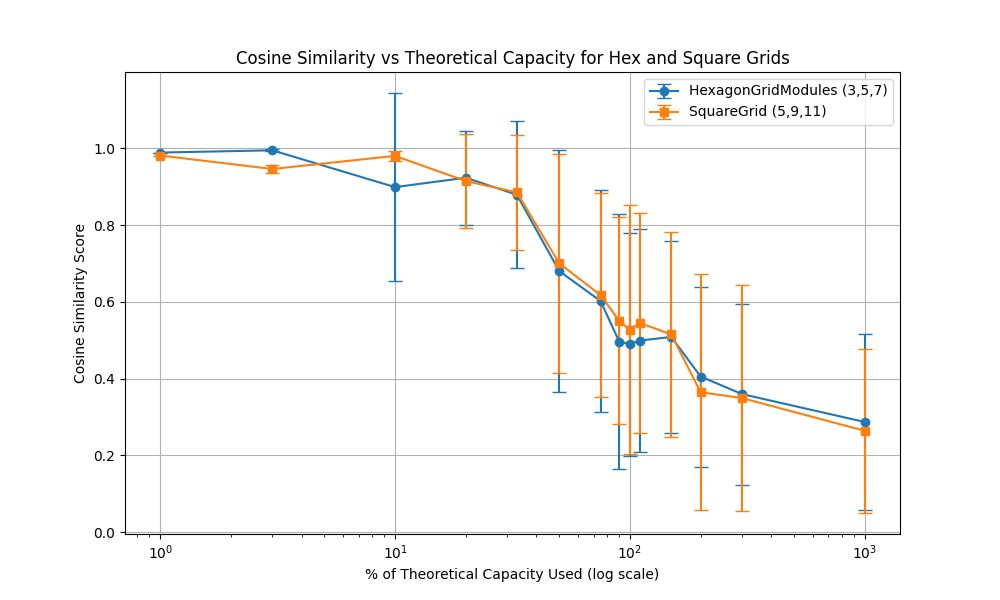}
        \caption{CIFAR Images}
        \label{fig:NumImageCIFAR}
    \end{subfigure}
    
    \caption{Average Cosine Similarity Scores vs \% of Theoretical Maximal Memory Capacity Filled for Different Image Datasets: MNIST, Fashion MNIST, and CIFAR.}
    \label{fig:NumImageComparison}
\end{figure}
 
 Across all three datasets we see that recall quality of images as a function of the percentage of capacity of the memory which is used is indistinguishable whether we used square grid cells or hexagonal grid cells (cf. Fig. \ref{fig:NumImageComparison})

\begin{figure}[!htbp]
    \centering
    \includegraphics[scale=0.07]{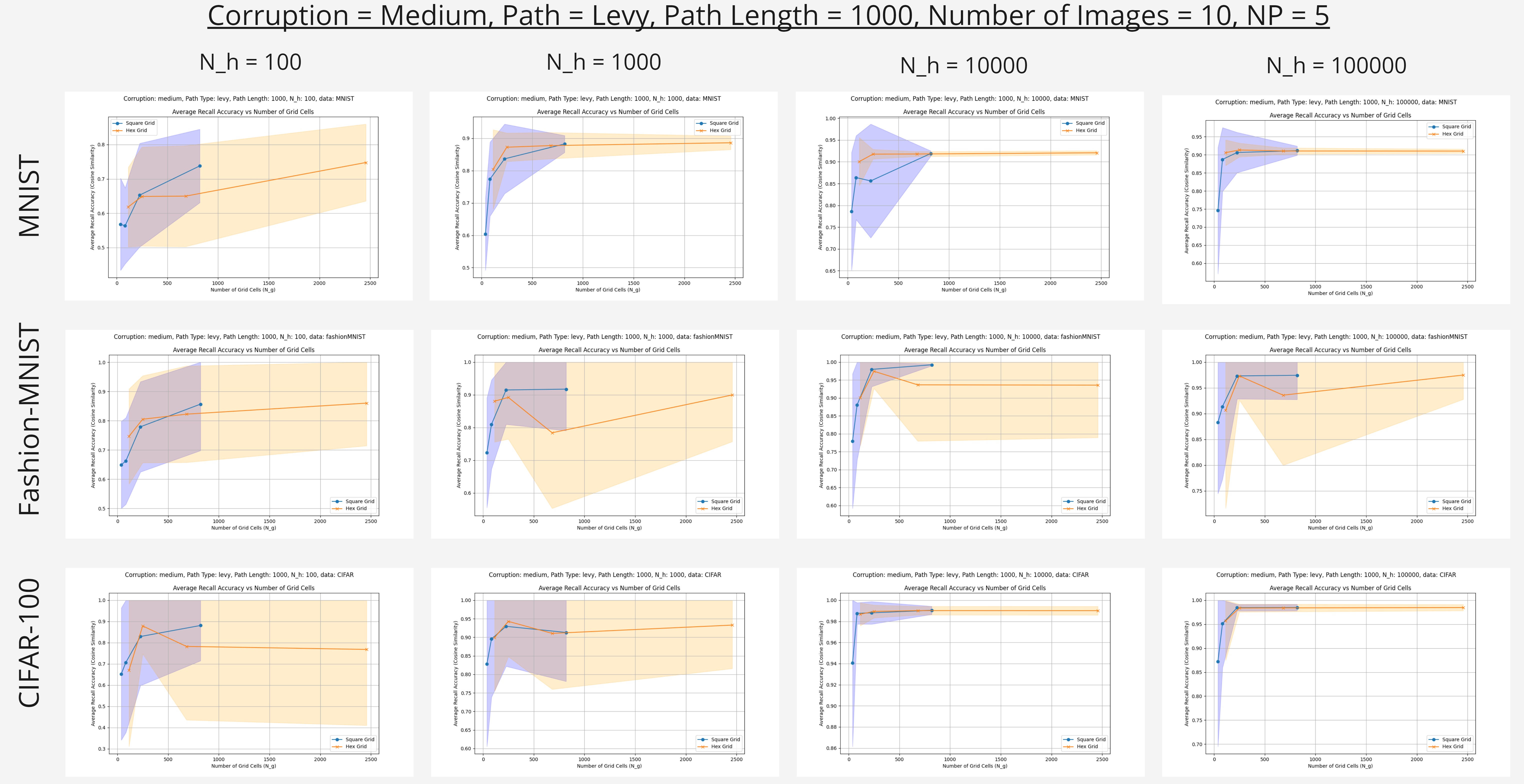}
    \caption{Cosine Similarity Recall Scores vs Number of grid cells, $N_g$, and hippocampal place cells, $N_h$.}
    \label{fig:varyNH}
\end{figure}

 \textbf{Experiment 2: When the number of $N_h$ is varied (Figure \ref{fig:varyNH}):} Looking across the three datasets (MNIST, Fashion-MNIST, and CIFAR-100), that for wide variation in the number of hippocampal place cells, $N_h$, and grid cells $N_g$, the storage and recall performance is identical for both square and hexagonal grid cell tilings.

 \textbf{Experiment 3: On different path types and path sizes:} Across all datasets MNIST (Figure: \ref{fig:MNISTPath}), FashionMNIST (Figure: \ref{fig:FashionMNISTPath}), and CIFAR100 (Figure: \ref{fig:CIFARPath}) the results reveal consistent patterns when varying the path types (Straight Line, Brownian, and Levy) and path lengths. A key observation is that both hexagonal and square grids exhibit similar performance across most path types and lengths, with minor deviations. 

\section{Conclusions}\label{conclusions}
In our experiments, we evaluated the performance of square and hexagonal grid cells within the Vector-HaSH model using various image datasets and path simulations. The results indicate that hexagonal grid cells display a similar performance to square grid cells across all our experiments, as measured by the mean cosine similarity and standard deviation in memory recall. 

Thus, in the context of computational models for memory recall, the biological benefits of hexagonal tiling do not necessarily translate into a computational advantage. The similar performance between hexagonal and square grid cells indicates that both structures are equally capable of encoding and recalling spatial information, at least within the tested scenarios. The underlying neural mechanisms in the brain might leverage hexagonal grids not for enhanced computational performance but due to the natural formation tendencies and structural efficiencies of biological tissues. This finding suggests that the brain's preference for hexagonal grids may be more about the ease of biological implementation.

\subsection{Future Directions}

These results open new avenues for investigating the role of grid cell structures in the brain beyond memory recall. Further research could explore other cognitive tasks where the hexagonal arrangement might offer a computational advantage or provide insights into how biological constraints shape neural coding strategies.

Another promising direction is to extend the testing of hexagonal grids to 3D environments, such as those provided by AnimalAI. This extension would allow us to evaluate whether the advantages of hexagonal grids observed in 2D spaces also apply to more complex, three-dimensional settings. Testing in 3D environments would provide a more comprehensive understanding of how grid structures operate in realistic scenarios, potentially uncovering new insights into the spatial processing capabilities of hexagonal grids in naturalistic tasks.

\section{Acknowledgments}
We would like to extend our heartfelt thanks to Tianyi Xu for his valuable advice and assistance during the implementation phase of Memory Storing \& Recalling, as well as for his contributions to the literature review. His efforts were crucial to the success of this project.

\bibliographystyle{plain}
\bibliography{reference}

\newpage
\appendix
\section{Supplementary Material}
\subsection{Experiment Setup}
Our variables consist of path type, path length, number of hippocampus cells (\(N_h\)), Number of images to store, grid module configurations, corruption level and we used a hyperparameter, $N_P = 5$ (See Section \ref{app: c_parameter}) which is the expected dimension of the projection from grid cells to place cells. We carried out several experiments with these variables on three different datasets to answer, How do square grids perform vs hexagonal grid cells: 
\begin{itemize}
    \item  \textbf{With varying amounts of stored images:} We stored images that would take up $x \in (1\%, 3\%, 10\%, 20\%, 33\%, 50\%, 75\%, 90\%, 100\%, 110\%, 150\%, 200\%, 300\%, 1000\%)$ of the theoretical memory capacity, where the theoretical memory capacity in number of images is:
    \[
    memory\_capacity = \frac{N_g * N_h}{\text{Number of floats in image}}
    \]
    for $N_h = 1000$, Hexagon Grid Module Configuration = $(3, 5, 6)$, Square Grid Module Configuration = $(5, 9, 11)$ and medium corruption.

    \item \textbf{When the number of $N_h$ is varied:} We set corruption = 'medium', path = 'levy walk', path length = 1000, number of images = 10, and compared recall performance of $N_h \in \{100, 1000, 10000, 100000\}$. For each $N_h$, we averaged the cosine similarity scores for $\text{grid\_module\_configurations} \in [[2, 3, 5], [3, 5, 7], [5, 9, 11], [11, 13, 23]]$ and across $k = 5$ random generated levy paths. We tested these on MNIST, Fashion-MNIST and CIFAR-100

    \item \textbf{On different path types and path sizes:} We set corruption = 'medium', $N_h = 10000$, number of images ($N$) = 10, and compared recall performance of $path\_length \in \{N, N^2, N^3, N^4\}$ and $path\_type \in \{\text{straight path, brownian, lévy}\}$. For each $path\_type$, $path\_type$ combination, we averaged the cosine similarity scores for $\text{grid\_module\_configurations} \in [[2, 3, 5], [3, 5, 7], [5, 9, 11], [11, 13, 23]]$ and across $k = 5$ random generated levy paths.
\end{itemize}

\subsection{Supplementary Result Plots} \label{Results}

\begin{figure}[!htbp]
    \centering
    \includegraphics[scale=0.07]{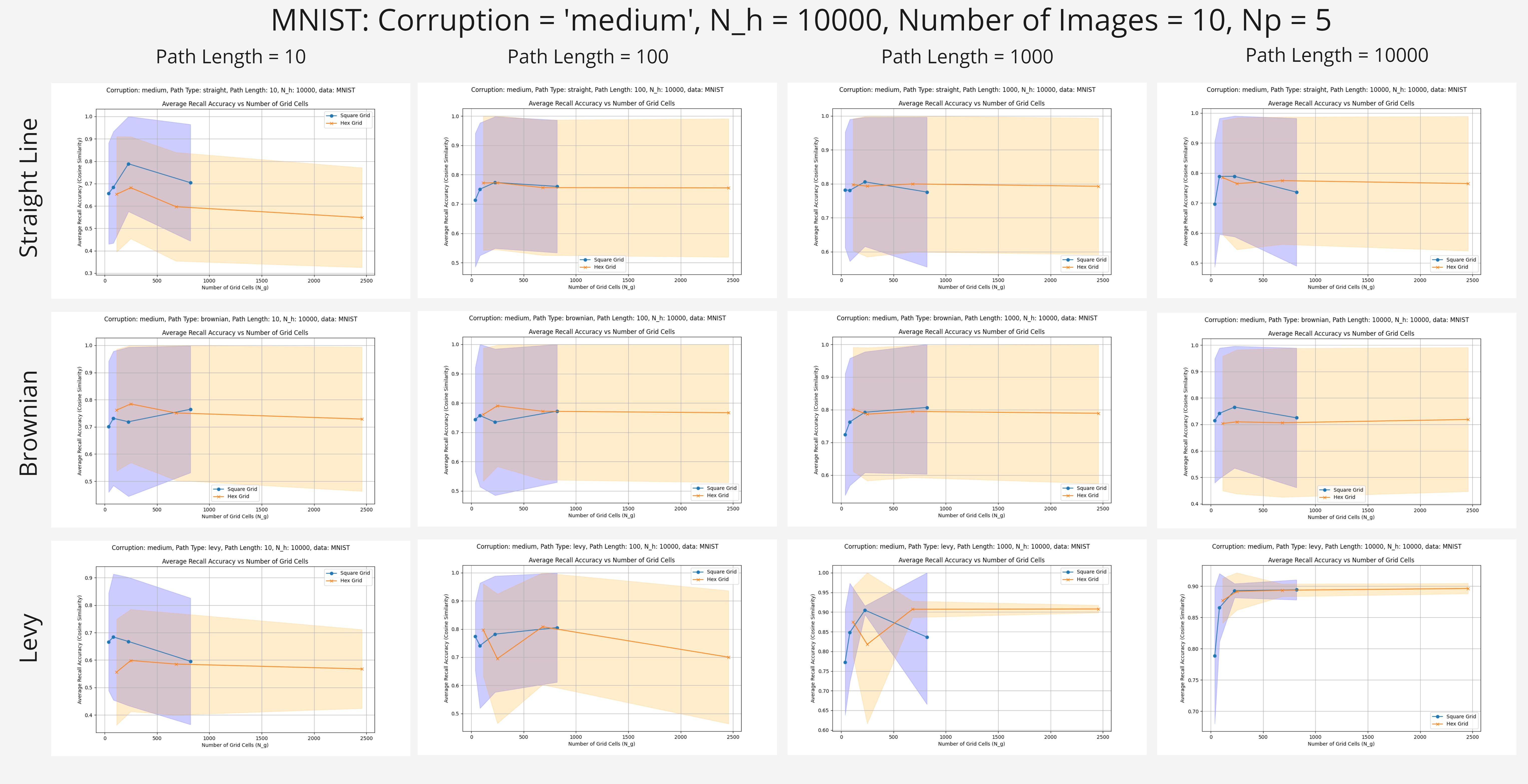}
    \caption{MNIST: Cosine Similarity Recall Scores vs Path Type and Path Length}
    \label{fig:MNISTPath}
\end{figure}
\begin{figure}[!htbp]
    \centering
    \includegraphics[scale=0.07]{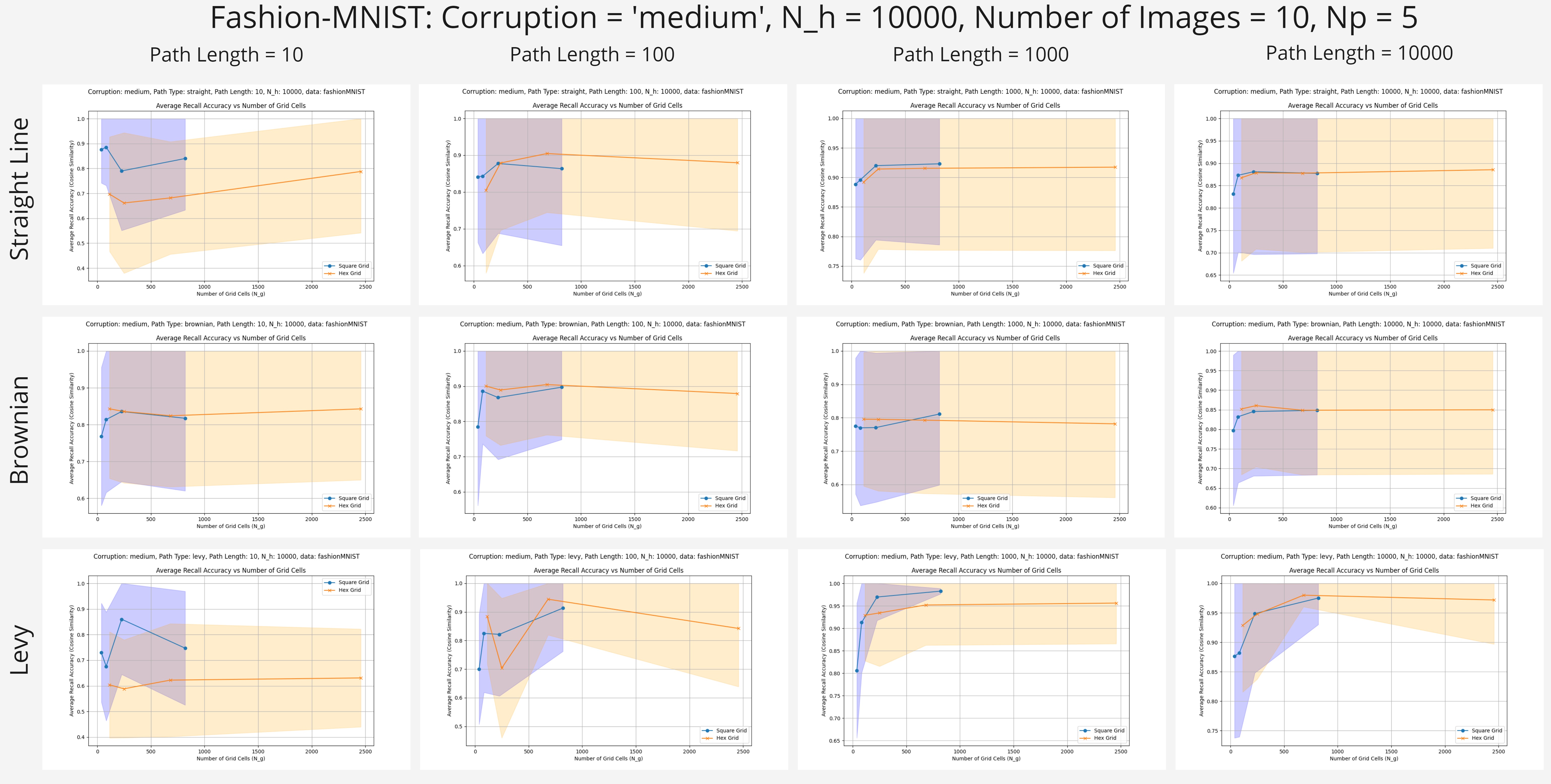}
    \caption{Fashion-MNIST: Cosine Similarity Recall Scores vs Path Type and Path Length}
    \label{fig:FashionMNISTPath}
\end{figure}
\begin{figure}[!htbp]
    \centering
    \includegraphics[scale=0.07]{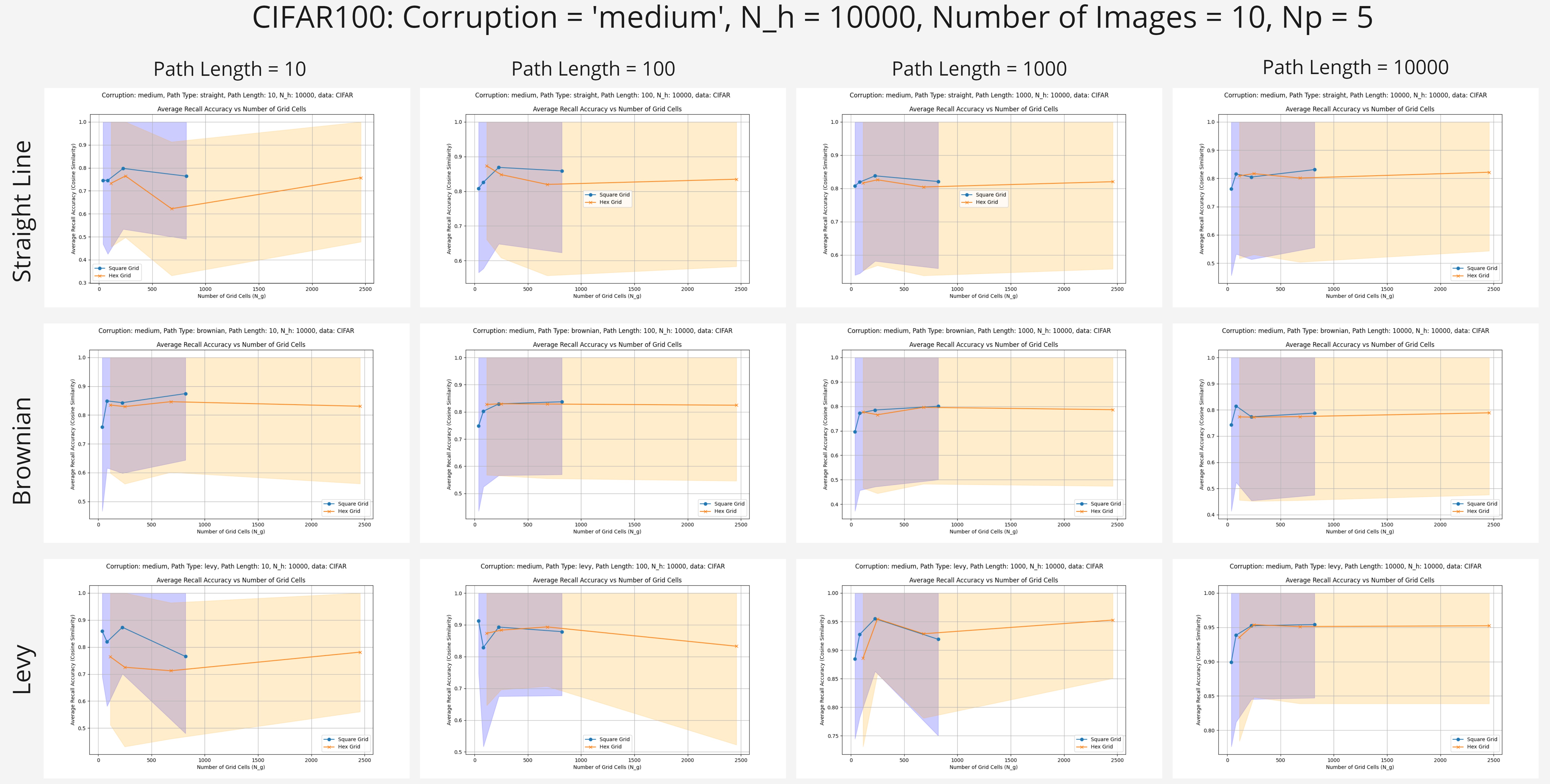}
    \caption{CIFAR: Cosine Similarity Recall Scores vs Path Type and Path Length}
    \label{fig:CIFARPath}
\end{figure}

\subsection{Methodology}\label{methods}
In this section, we detail the implementation and evaluation of square and hexagonal grid cells within the Vector-HaSH model, describing the processes of memory storage and recall, and the benchmarks used to assess their performance.

\subsubsection{Square Grid-Based Spatial Representation}
Square grid cells serve as the foundational structure in our implementation of the Vector-HaSH model.

\begin{enumerate}
    \item \textbf{Grid Representation:} A single square grid is represented by two primary one-hot encoded vectors, \(g_x\) and \(g_y\), corresponding to the $x$ and $y$ coordinates in a 2D square grid. Each vector has a length $\lambda$ defining the length of a side of the square grid. 

    \item \textbf{Grid Vector Formation:} The overall grid vector \(g\) is formed by computing an outerproduct of \(g_x\) and \(g_y\) as in Figure \ref{fig:outer_product_matrix}:
    
    \begin{figure}[!htbp]
        \[
            \texttt{square\_grid = np.outer(g\_x, g\_y)}
        \]
        \begin{center}
            \includegraphics[scale=0.2]{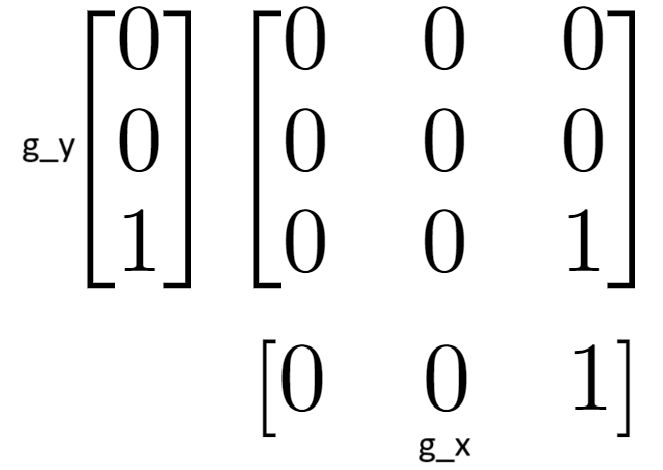}
        \end{center}
        \caption{The outer product of vectors \( \mathbf{g_y} \) (left) and \( \mathbf{g_x} \) (bottom), generating a matrix representation of the grid. The vector \( \mathbf{g_y} \) defines the rows, while \( \mathbf{g_x} \) defines the columns.}
        \label{fig:outer_product_matrix}
    \end{figure}

    \item \textbf{Cartesian to Grid Coordinate:} We can achieve the conversion from Cartesian to grid coordinate by simply applying the modulus operation to the $x$ coordinate and the $y$ coordinate with the \textit{lambda} for that grid module:
    \[
        gx_{activated\_index} = x \hspace{0.1cm} \% \hspace{0.1cm} \lambda
    \]
    \[
        gy_{activated\_index} = y \hspace{0.1cm} \% \hspace{0.1cm}\lambda
    \]
    This will return a number that serves as the index indicating which digit should be set to 1 in $g_x$ or $g_y$ to represent the Cartesian coordinate. Therefore, given a coordinate $(x, y)$, we can map this onto a squared grid using our representation. Here, we define the complete matrix as a \textit{'Grid Module'} and each digit in the module as a \textit{'Grid Cell'}.

    \item \textbf{Modular Structure:} The structure of our complete grid is modular. That is, our complete grid consists of multiple grid modules of different sizes defined by an array of \textit{lambdas}. Then, the total size for a grid module is:
    
    \[
    \text{Module Size} = \lambda^2
    \]
    and the size of the full grid is:

    \[
    \text{Grid Size} = \sum_{\text{all }\lambda} \lambda^2
    \]
    
    This modular approach aids in handling larger grids by dividing the grid into manageable parts, where each module is processed independently during operations such as denoising or updating. In addition, the combination of these grid modules allows for more spaces in the environment to be mapped by our grid, therefore allowing the representation of a larger range of (x, y) coordinates without overlap.

    \item \textbf{Grid Vector:} For the final grid vector, we flatten each grid module into a vector and concatenate the flattened modules into a single vector $g$.
    \begin{figure}[!htbp]
        \begin{center}
            \includegraphics[scale=0.45]{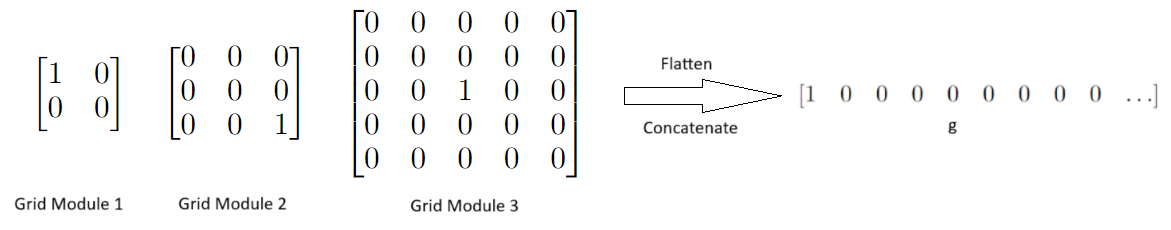}
        \end{center}
        \caption{Flattening and concatenating grid modules 1, 2, and 3 into a unified vector.}
        \label{fig:grid_flatten_concatenate}
    \end{figure}    
\end{enumerate}

\subsubsection{Hexagonal Grid-Based Spatial Representation}
Hexagonal grid cells are implemented as an alternative spatial representation aiming to leverage the natural efficiency of hexagonal tiling for spatial tasks. Unlike square grids, hexagonal grids are closer to the spatial encoding observed in biological systems. In our implementation, a single hexagon is used to represent the entire hexagonal grid.

\begin{enumerate}
    \item \textbf{Hexagonal Tiling and Grid Modules:} The hexagonal grid is composed of a collection of hexagonal grid modules, each defined by the size parameter \(R\) (or equivalently $\lambda$ as for the square grid modules), which represents the distance from the center of a hexagon to one of its corners. Each grid module is represented as a 'big' hexagon with smaller hexagons (grid cells) within it. 'Big' hexagons have $R > 1$ while small hexagons have $R = 1$.
    
    These hexagons are arranged in layers, with each layer representing a ring of hexagons around a central origin. This tiling method ensures that the grid covers the space uniformly, providing a natural representation of spatial locations. The size of each hexagonal grid module is given by:
    
    \[
    \text{Module Size} = 3 \times \lambda^2
    \]
    where \(\lambda\) is the radius of the hexagon (figure \ref{fig:Hexagonal tiling and grid modules}, Proof: \ref{app:3R2Proof}).

    \begin{figure}[!htbp]
        \centering
        \includegraphics[width=0.4\linewidth]{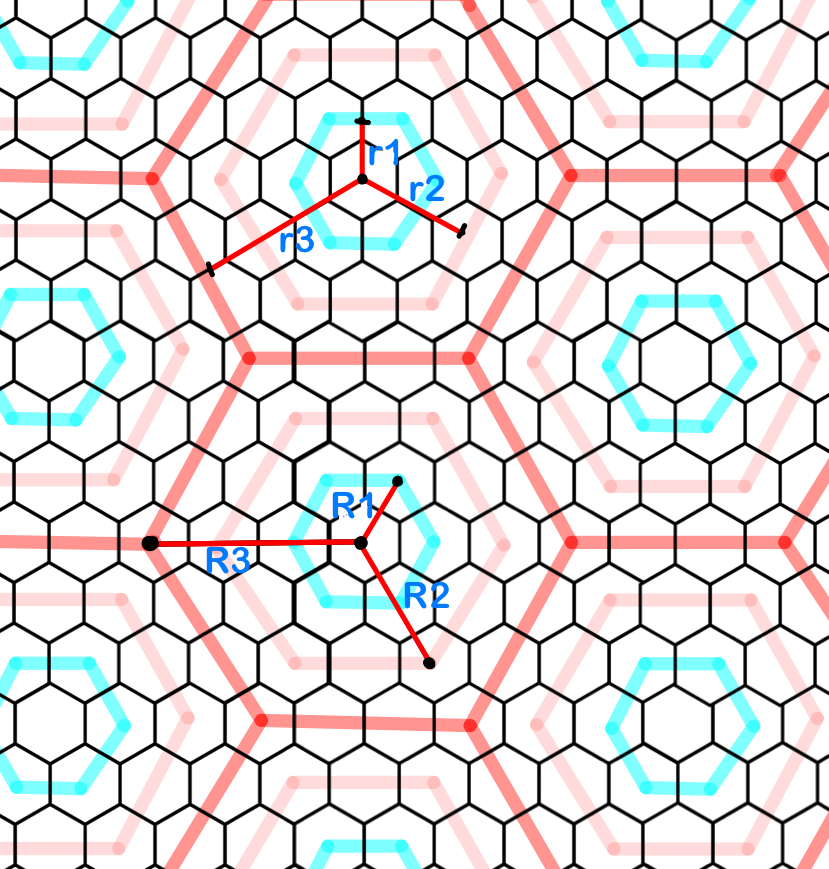}
        \caption{Illustration of our hexagonal tiling structure. For the hexagons grid modules, each radius \(R\) corresponds to a specific grid module size parameter \(\lambda\), with examples shown for \(R_1 = \lambda_1 < R_2 = \lambda_2 < R_3 = \lambda_3\). The distance from the origin to the midpoint of a hexagon’s side is given by \(r = \frac{\sqrt{3}}{2} \cdot R\). For the grid cells (small hexagons in black), we have $R = 1$.}
        \label{fig:Hexagonal tiling and grid modules}
    \end{figure}

    \item \textbf{Assignment of IDs:}  To identify which grid cell is activated, each hexagon cell in the grid is assigned a unique identifier (ID). The ID assignment begins at the central hexagon and proceeds outward in layers. Special care is taken to ensure that hexagons cells that are in locations mimicking a corner of an equilateral triangle share the same ID. as well as opposite hexagons, to maintain symmetry and consistency across the grid (See  \ref{app:3R2Proof}) (Figure \ref{fig:Assignment of IDs in the grid structure}).
    
   \begin{figure}[!htbp]
        \centering
        \includegraphics[width=0.4\linewidth]{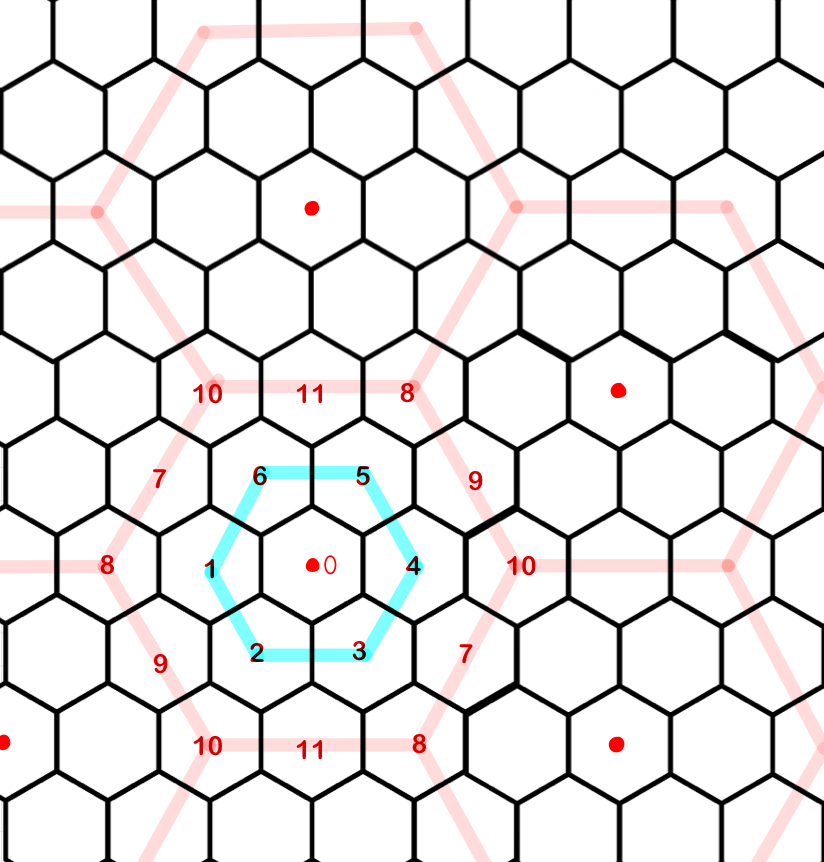}
        \caption{Illustration of ID assignment in the hexagonal grid structure.}
        \label{fig:Assignment of IDs in the grid structure}
    \end{figure}

    \item \textbf{Coordinate Transformation:} Given that the hexagonal grid does not align perfectly with Cartesian coordinates, a transformation is necessary to convert between the original Cartesian coordinates \((x, y)\) and the hexagonal grid coordinates \((x', y')\). First we define certain properties of each hexagon (Figure: \ref{fig:hexfeatures}):
    \begin{figure}[!htbp]
        \centering
        \includegraphics[scale=0.3]{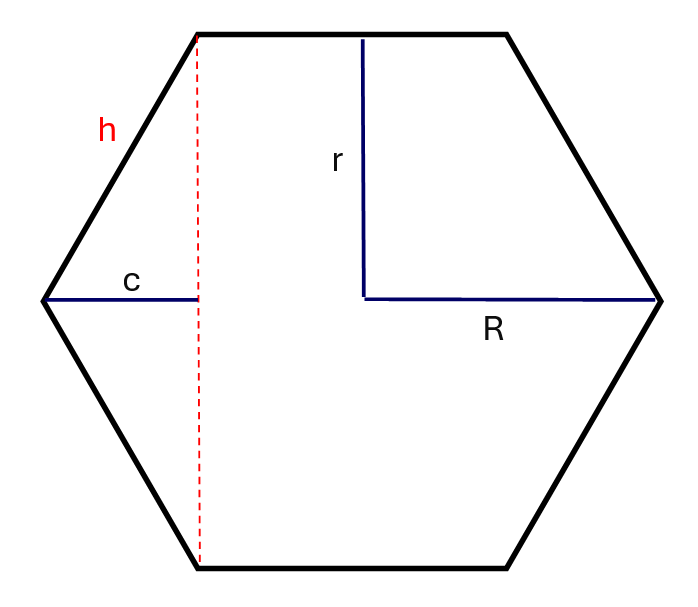}
        \caption{Geometrical features of a hexagonal grid cell (or module). $R$ denotes the distance from the center of the hexagon to its vertex (corner), while $r$ represents the distance from the center to the midpoint of a side. The variable $c$ defines the distance from the corner to the highlighted red line, as depicted in the figure.}
        \label{fig:hexfeatures}
    \end{figure}

    \[
    r = \frac{\sqrt{3}}{2} R,
    \]
    Note that we assume the use of regular hexagons. Therefore, the gradient of the line $h$ is $\sqrt{3}$ such that :
    \[
    gradient_h = \sqrt{3} = \frac{y_2 - y_1}{x_2 - x_1} = \frac{r}{c},
    \]
    Rearranging, we get:
    \[
    c = \frac{1}{\sqrt{3}} \cdot r = \frac{1}{\sqrt{3}} \cdot \frac{{\sqrt{3}}}{2}R = \frac{1}{2}R
    \]
    Consider the small hexagon cells in Figure \ref{fig:subfig1}, note that these are not aligned at the Cartesian origin. The hexagon grid modules don't start at (0, 0) on the Cartesian axis (Figure \ref{fig:subfigure_7}). As a result, we apply a transformation to the Cartesian coordinates $(x', y')$ to map them onto the hexagonal grid to get $(x, y)$:
    \[
    x =  x' + \frac{R}{2}, \hspace{0.1cm} y = y' + r.
    \]
    where \((x, y)\) are the transformed coordinates in the hexagonal grid's system.

    To reverse the transformation and convert from the hexagonal grid's coordinates \((x, y)\) back to the original coordinates \((x', y')\), the following expressions are used:
    
    \[
    x' = x - \frac{R}{2}, \hspace{0.2cm} y' = y - \frac{R}{2}.
    \label{eq:coordinate_transform}\]

    Note that for hexagon grid modules, they are overlayed over the grid of smaller hexagon grid cells. Operations on this overlayed grid are the same with a couple differences:
    \begin{itemize}
        \item The $x$ and $y$ coordinates are inverted as the grid modules are rotated $90^{\circ}$.
        \item The grid modules have $R > 1$.
    \end{itemize}

    These transformations ensure that the coordinates used for grid operations maintain spatial consistency with the original Cartesian system (Figure \ref{fig:subfigure_7}).

    \begin{figure}[h]
        \centering
        \begin{subfigure}[t]{0.3\linewidth}
            \includegraphics[width=\linewidth]{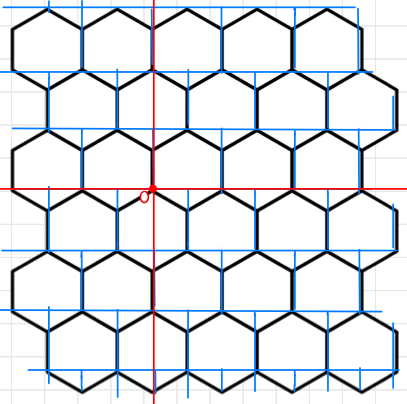}
            \caption{Initial square grid tiling in the Cartesian coordinate system.}
            \label{fig:subfig1}
        \end{subfigure}
        \begin{subfigure}[t]{0.3\linewidth}
            \includegraphics[width=\linewidth]{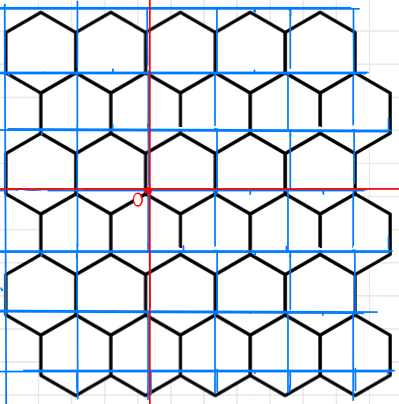}
            \caption{Modified tiling after shifting the odd-numbered rows leftward.}
            \label{fig:subfig2}
        \end{subfigure}
        \begin{subfigure}[t]{0.3\linewidth}
            \includegraphics[width=\linewidth]{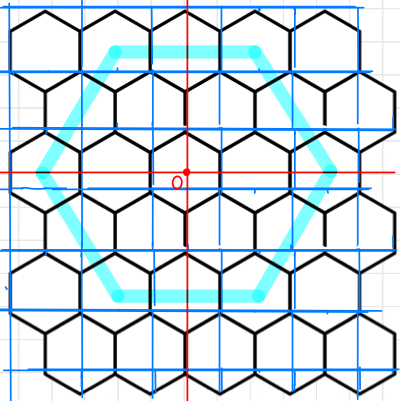}
            \caption{Final configuration with the origin aligned at the center of the hexagonal grid cell.}
            \label{fig:subfig3}
        \end{subfigure}
        \caption{Transformation sequence illustrating the transition from Cartesian coordinates to the appropriate hexagonal grid location.}
        \label{fig:subfigure_7}
    \end{figure}
    
    \item \textbf{Cartesian to Hex Coordinates:} We first consider the Hexagon Grid modules and assume we are in the space described by Figure \ref{fig:subfig3}, with the hexagons (small and big) centered at $(0, 0)$. Given a Cartesian coordinate \((x, y)\) in that infinite space, we need to be able to represent this point using a single hexagon (finite space representing an infinite grid).
    \\
    \\
   To accomplish this, we need to identify the midpoint of the hexagon grid module containing \((x, y)\) in that subspace. The first step in this process is to apply a transformation to center the hexagons from $(0, 0)$ to ($\frac{r}{2}$, $R - c$) for simpler calculations. Also note that hexagon grid modules are rotated as compared to hexagon cells, so we interchange the $x$ and $y$ during transformation to get \((x', y')\). We then overlay another grid of rectangles over the grid of hexagons modules (Figure: \ref{fig:subfig1}) and determine which rectangle \((x', y')\) is in. Note that the height of the rectangle is $2R - c$, the width is $2r$ and the rectangular grid is centered with it's bottom left corner at $(0, 0)$ in the Cartesian System (Figure: \ref{fig:subfig3}). Next, with the given $(x', y')$ coordinates and the rectangle measurements, we determine which rectangle the point lies in.
    \\
    \\
    With this information, the point can be in one of three hexagons (Figure: \ref{fig:WhichHex}). We can determine the hexagon by calculating the midpoints of the hexagons and determining which hexagon's midpoint is closest to our coordinate $(x', y')$. From this we find our hexagon $h^*$.
    \begin{figure}[!htbp]
        \centering
        \includegraphics[width=0.4\linewidth]{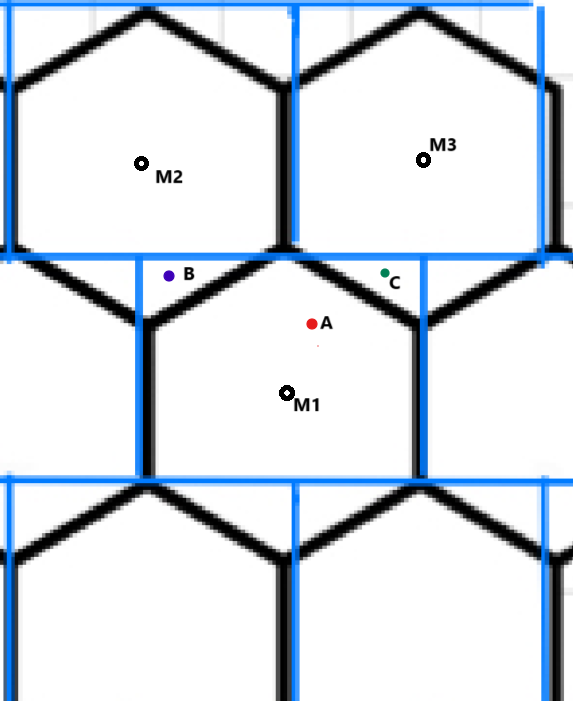}
        \caption{Consider the rectangle containing M1. Points A, B, C are in different hexagons but the same rectangle. We can use the distance from the hexagon midpoints to determine which hexagon these points lie in.}
        \label{fig:WhichHex}
    \end{figure}
    \\
    \\
    The next step is to find which hexagon cell in the hexagon grid module \((x', y')\) lies in. Note that in this case, we again start at the same point as we did with hexagon grid modules, except with $R = 1$ and the cells are not rotated unlike the hexagon grid modules, so we don't interchange the $x, y$ coordinates.
    \\
    \\
    Using the initial $(x, y)$ coordinate, we apply the same procedure (without interchanging the $x$, $y$ coordinate) to obtain the location of this point in the small hex cell coordinate system. With this we find the small hexagon $(x, y)$ is in. Let this be $h^*_{small}$.
    \\
    \\
    We can then subtract the midpoints of $h^*_{small}$ and $h^*$ to obtain the relative position of the small cell with respect to the midpoint of $h^*$. Then using our hexagon cell numbering system, we can assign an ID to that cell. Telling us which cell is activated in that hexagon grid module. We then create a vector of size $3R^2$, and assign the index equal to ID to be 1. This process is repeated for each grid module and the vectors for each module are concatenated.
    \end{enumerate}





        


\subsubsection{Memory Encoding \& Storage}
\begin{figure}[!htbp]
        \centering
        \begin{subfigure}[b]{0.45\linewidth}
            \centering
            \includegraphics[width=\linewidth]{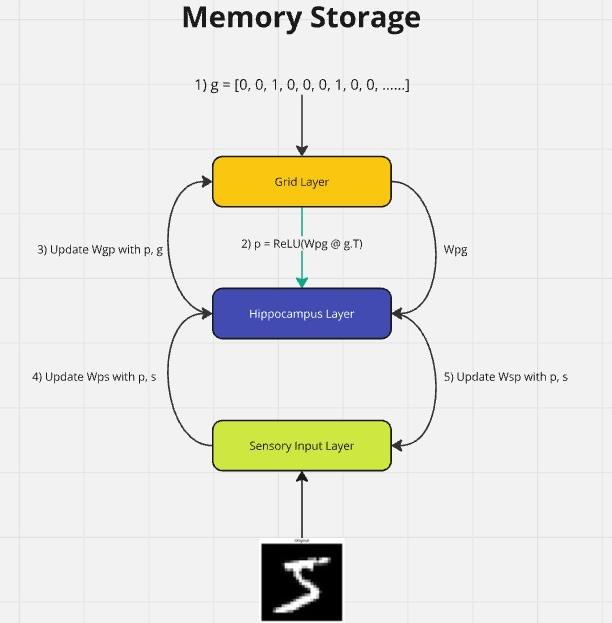}
            \caption{Memory storage process in the Vector-HaSH framework.}
            \label{fig:store_memory}
        \end{subfigure}%
        \hfill
        \begin{subfigure}[b]{0.45\linewidth}
            \centering
            \includegraphics[width=\linewidth]{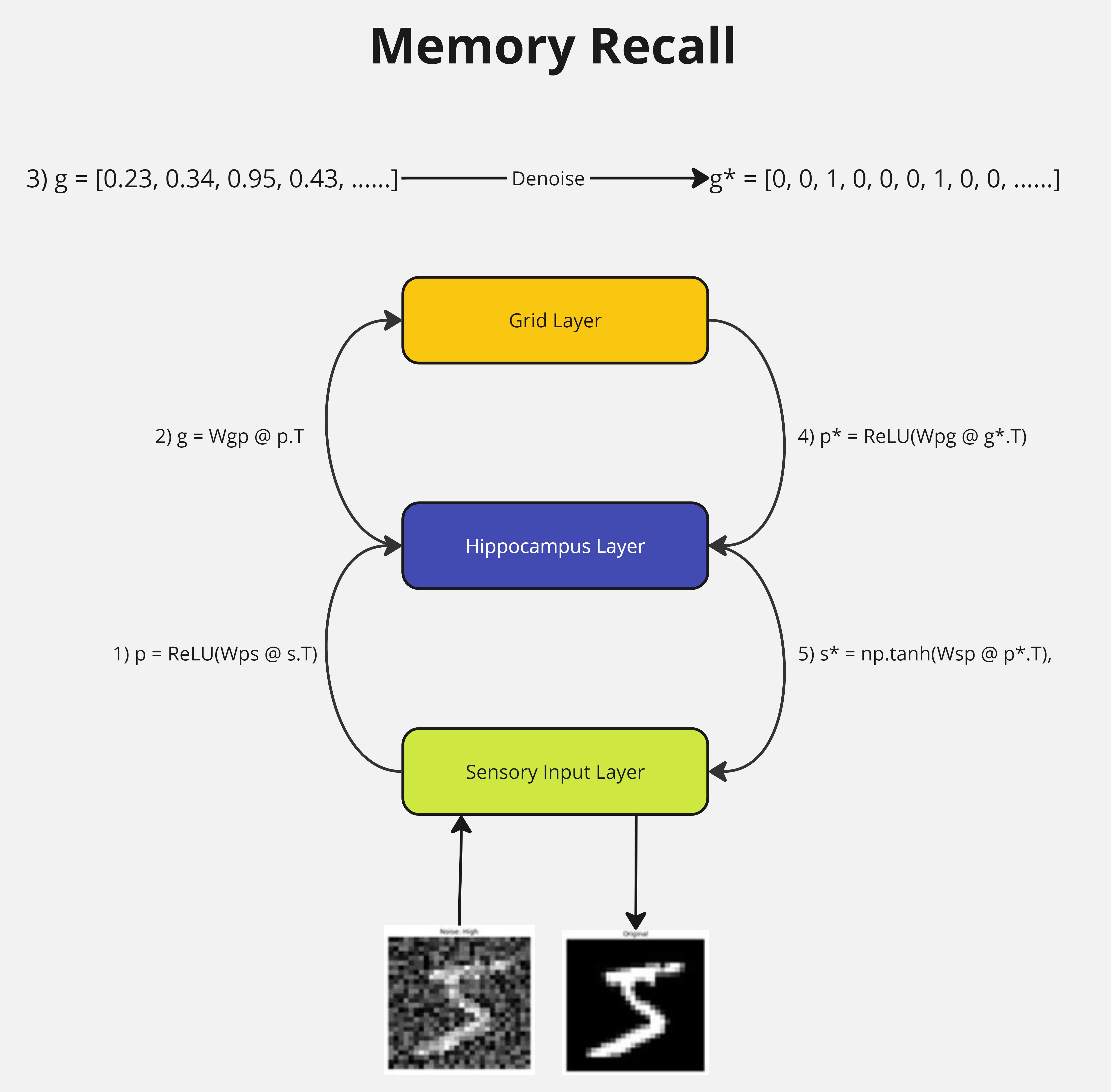}
            \caption{The process for recalling memories in Vector-HaSH.}
            \label{fig:recall_memory}
        \end{subfigure}
        \caption{Memory storage and recall in Vector-HaSH.}
        \label{fig:memory_process}
    \end{figure}

Memory storage in the Vector-HaSH model involves the following steps (Figure: \ref{fig:store_memory}):

\begin{enumerate}
    \item \textbf{Initialization of Weights:} The memory model initializes several weight matrices that connect the grid cells (\(g\)), hippocampal place cells (\(p\)), and sensory inputs (\(s\)). These connections form two crucial components of the model: the \textit{scaffold} and \textit{heteroassociation} mechanisms. The weights include \(W_{pg}\) (grid to place cells), \(W_{gp}\) (place to grid cells). These two weight matrices make up the 'memory scaffold', which is how the model stores it's memories. The other weight matrices are: \(W_{ps}\) (sensory to place cells), and \(W_{sp}\) (place to sensory cells). These matrices provide the hetero-association of the sensory input to the stored memory in the scaffold.
    
    \(W_{pg}\) is initialized using a normal distribution, with hyperparameters \(c\) (Section:\ref{app: c_parameter}) and \(var\) determining the mean and variance, respectively. After initialization, \(W_{pg}\) is fixed and does not change during the learning process, while the other weights are initialized to $0$ and are updated as memories are stored. 
    
    \item \textbf{Memory Storage:} First, given a coordinate, we determine the state of the grid vector $g$. With this, we can compute the input into the hippocampus layer, $p$, using the fixed matrix of random projections from grid cells to hippocampus cells $W_{pg}$:
    \begin{align}
    p = \text{ReLU}(W_{pg} \cdot g)
    \label{eq:g2p}
    \end{align}
    Once we have both the grid vector $g$ and hippocampus place cell activation $p$, we can begin constructing the scaffold by updating the weights:
    \begin{align}
    W_{gp} \leftarrow W_{gp} + \frac{g \cdot p^T}{\|p\|^2 + \epsilon}
    \label{eq:learnWgp}
    \end{align}
    Note that we can use the sensory input $s$, along with the newly calculated $p$ to update the weights that handle hetero-association of the sensory input to the location in the scaffold:
    \begin{align}
    W_{sp} \leftarrow W_{sp} + \frac{s \cdot p^T}{\|p\|^2 + \epsilon}
    \label{eq:learnWsp}
    \end{align}
    \begin{align}
    W_{ps} \leftarrow W_{ps} + \frac{p \cdot s^T}{\|s\|^2 + \epsilon}
    \label{eq:learnWps}
    \end{align}
    The update process follows a Hebbian-like learning rule, where the weight matrices \(W_{gp}\), \(W_{sp}\) and \(W_{ps}\) are incrementally updated using the following expressions:

    These updates ensure that the weights are adjusted based on the correlation between the grid cells, place cells, and sensory inputs, with normalization to maintain stability in the learning process. 
\end{enumerate}

\subsubsection{Memory Retrieval \& Recall}
Memory recall is the process of retrieving stored memories from noisy or incomplete sensory inputs (Figure: \ref{fig:recall_memory}):

\begin{enumerate}
    \item \textbf{Noisy Input Processing:} The noisy sensory input $\tilde{s}$ is processed through the trained model to, first, compute the noisy place cell activation \(\tilde{p}\) using the learned matrix $W_{ps}$: 
    \begin{align}
        \tilde{p} = ReLU(W_{ps}  \cdot \tilde{s})
        \label{eq:noisyp}
    \end{align}
    The noisy place cell activation is then used to infer the corresponding noisy grid state \(\tilde{g}\) by applying the learned weights \(W_{gp}\):
    \begin{align}
        \tilde{g} = W_{gp} \cdot \tilde{p}
        \label{eq:noisyg}
    \end{align}

    \item \textbf{Denoising and Modular Reconstruction:} The inferred grid state is denoised using the modular grid structure through a "winner-takes-all" (WTA) mechanism, similar to processes observed in the human brain. This mechanism involves splitting the noisy grid into modules, selecting the neuron or cell with the maximum activation within each module, and suppressing the others. The result is a reconstructed clean grid state, $g^*$, where only the most strongly activated cells remain active. This denoising process ensures that the grid state returns to a valid and stable representation within the grid space, effectively filtering out noise and enhancing the accuracy of the grid's spatial representation.

    \item \textbf{Sensory Input Reconstruction:} The denoised grid state is used to compute the place cell activation using the fixed random projection $W_{pg}$:
    \begin{align}
        p^* = ReLU(W_{pg}  \cdot g^*)
        \label{eq:denoisedp}
    \end{align}
    which is then mapped back to the sensory layer using the learned weight matrix \(W_{sp}\). The output is passed through a non-linear activation function (such as ReLU or tanh) to produce the final reconstructed sensory input.
    \begin{align}
    s^* = np.tanh(W_{sp} \cdot p^*)
    \label{eq:denoiseds}
    \end{align}
\end{enumerate}

\subsection{Proof: Size of a Hexagonal Grid Cell = \( 3r^2 \)} \label{app:3R2Proof}
Consider a hexagonal grid centered at a point, with a radius \( r \), which represents the maximum number of tiles away from the center. The goal is to determine the total number of grid cells, \( N_g \), within this hexagon.

\begin{enumerate}
    \item \textbf{Counting Interior Cells}:
    \begin{itemize}
        \item At distance \( r = 0 \), there is 1 cell (the red dot).
        \item At distance \( r = 1 \), there are 6 cells.
        \item At distance \( r = 2 \), there are 12 cells.
        \item In general, for distance \( r \), there are \( 6r \) cells.
    \end{itemize}
    Therefore, the total number of interior cells is the sum of the cells at each distance, for now we will ignore the outer border:
    
    \[
    N_{\text{interior}} = 1 + \sum_{j=1}^{r-1} 6j
    \]
    
    \item \textbf{Simplifying the Sum}:
    
    The sum of the series \( \sum_{j=1}^{r-1} 6j \) is given by:
    
    \[
    \sum_{j=1}^{r-1} 6j = 6 \times \frac{(r-1)r}{2} = 3r(r-1)
    \]
    
    Therefore, the total number of interior cells becomes:
    
    \[
    N_{\text{interior}} = 1 + 3r(r-1)
    \]
    
    \item \textbf{Counting Border Cells}:
    
    For the border cells, In a regular hexagonal grid, each hexagon shares its corner vertices with three other hexagons. More visually, consider the point in figure \ref{fig:overlap}:
    \begin{figure} 
        \centering
        \includegraphics[scale=0.3]{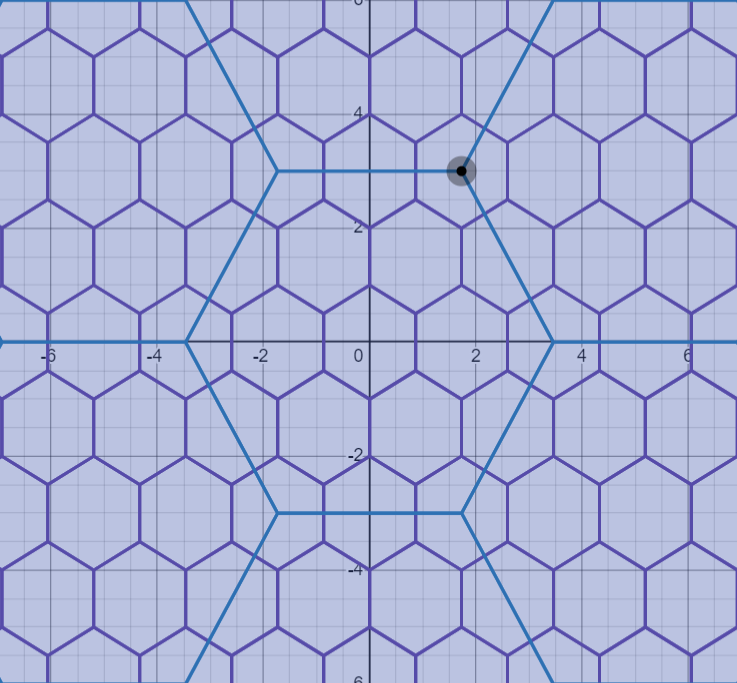}
        \caption{The black point can be a part of any three hexagons that are adjacent to it.}
        \label{fig:overlap}
    \end{figure}
    
    If we map the entire hexagonal grid in a single hexagon, we can represent this point in three ways, see Figure \ref{fig:overlap_rep}:
    \begin{figure} 
        \centering
        \includegraphics[scale=0.3]{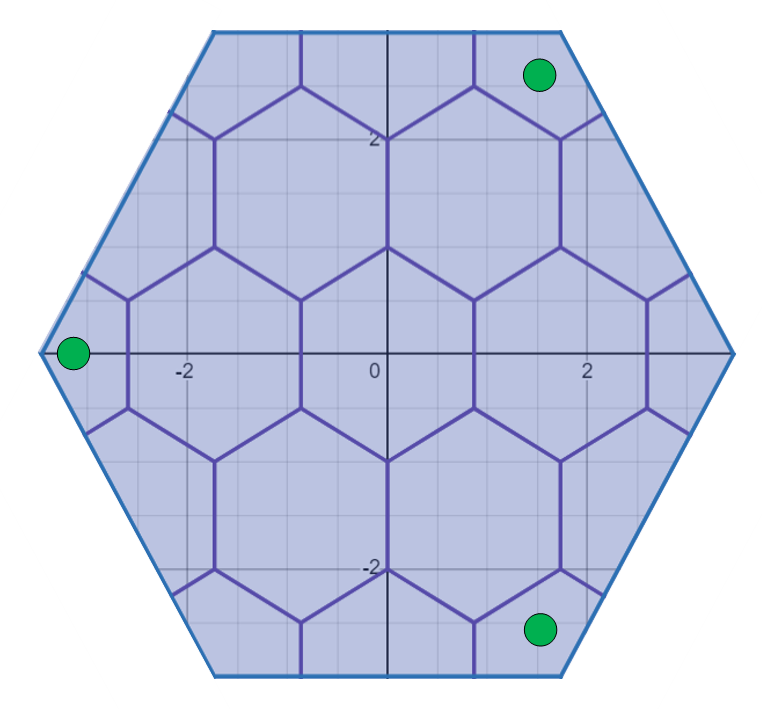}
        \caption{Representing the grid cell activated by the black point can be done in three ways.}
        \label{fig:overlap_rep}
    \end{figure}
    
    Therefore:
    
    \begin{itemize}
        \item There are 6 corners in a hexagon.
        \item Each corner is shared between three hexagons, meaning that each corner is counted multiple times when computing the border cells.
        \item Additionally, there are non-corner cells along the border, which are counted only twice, as they are shared between two adjacent hexagons.
    \end{itemize}
    
    The total number of border cells, accounting for the redundancy in counting corner cells and non-corner cells, is given by:
    
    \[
    N_{\text{border}} = 3(r-1)
    \]
    
    \item \textbf{Total Number of Cells}:
    
    Adding the number of interior cells and border cells, we have:
    
    \[
    N_g = 1 + 3r(r-1) + 3(r-1)
    \]
    
    After simplifying, we get:
    
    \[
    N_g = 3r^2
    \]
    
\end{enumerate}

Thus, the total number of grid cells in a hexagonal grid with radius \( r \) is \( 3r^2 \).

    
    

\subsubsection{Parameter Test for $N_p$} \label{app: c_parameter}

One key hyperparameter for the model is $N_P$ which represented the expected dimension of the hippocampal subspace in which grid cells are project through the random projection of equation \ref{eq:g2p}. We can set $N_P$ by tuning the mean and variance of the random Gaussian using to set matrix $W_{pg}$. We find, (see figures below) that the smaller $N_P$ is, the better the memory. But we don't want $N_P$ to be too small either as otherwise we risk projecting to a zero dimensional subspace (i.e. the zero vector) in some unlucky circumstances. We thus decided to set $N_P = 5$ to provide a balance between risk minimization and performance.

\begin{figure}[h!]
\centering
        \includegraphics[width=\textwidth]{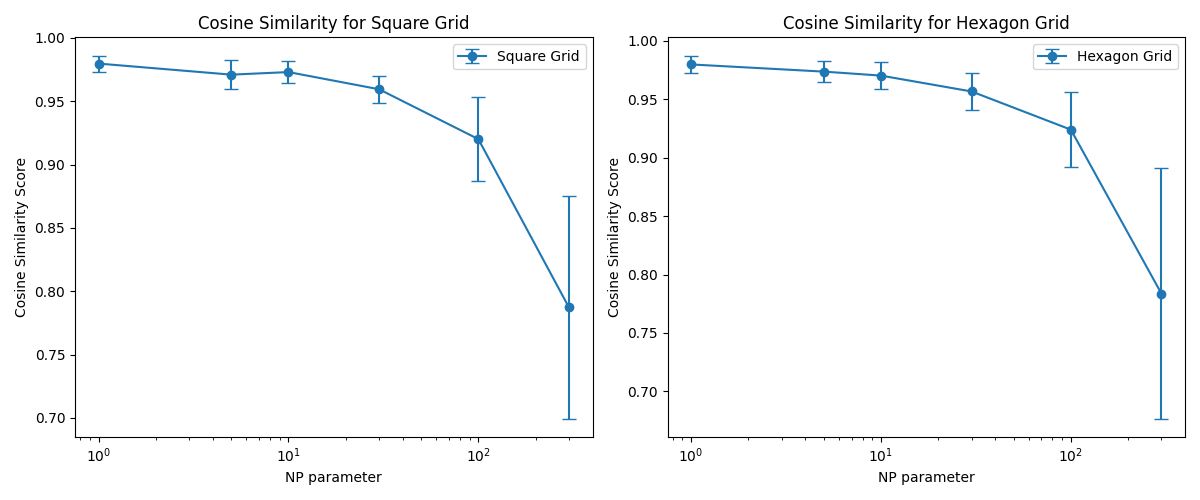}

    \caption{$N_P$ parameter test with hyperparameters:  Number of Images = 10, $N_h=1000$.}
    \label{fig:np-testing10images}
\end{figure}

\begin{figure}[h!]
    \centering
\begin{subfigure}{0.4\textwidth}
        \centering
        \includegraphics[width=\textwidth]{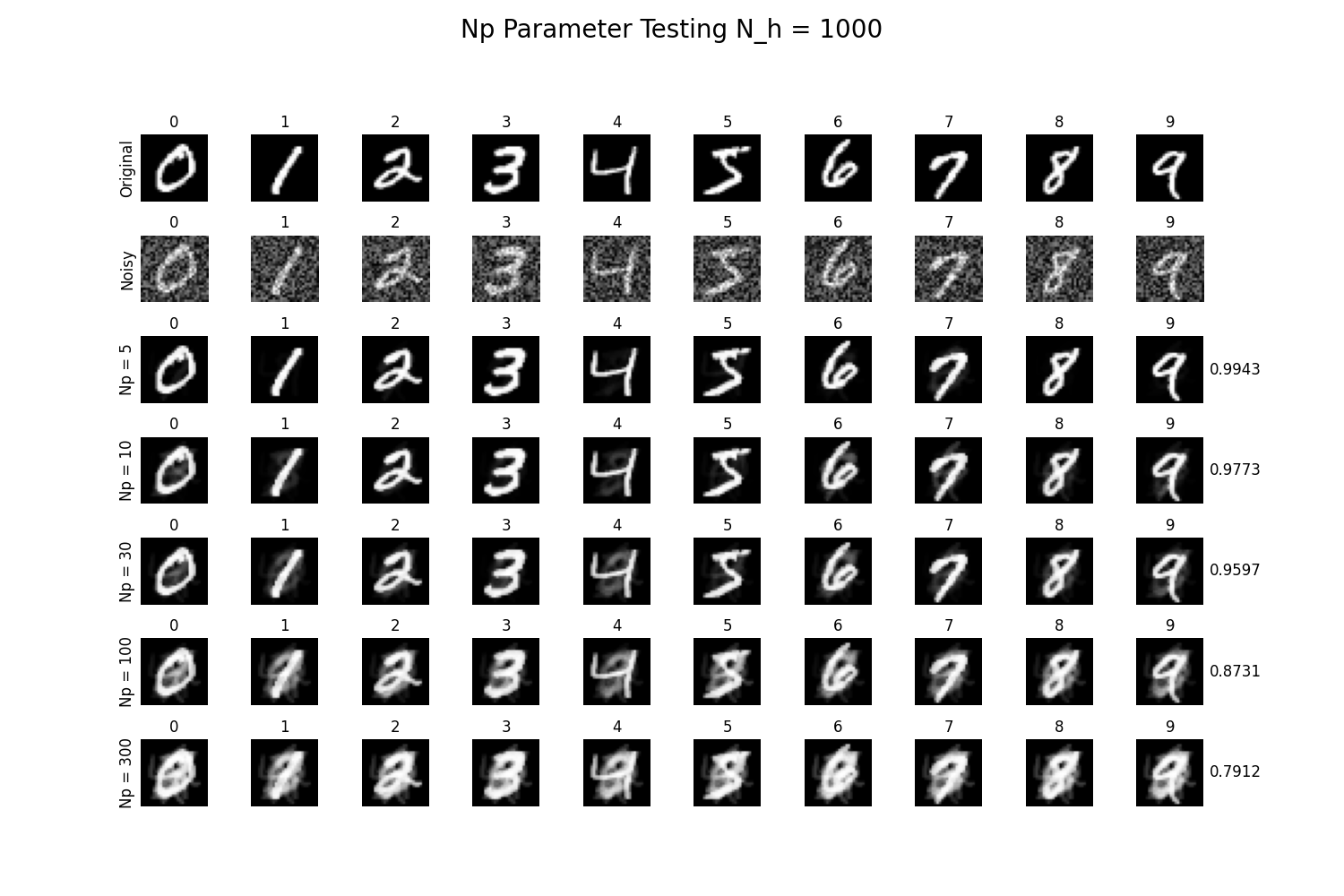}
        \caption{Recall Performance with $N_h$ = 1000.}
        \label{p_test_1000}
    \end{subfigure}
    \hfill
    \begin{subfigure}{0.4\textwidth}
        \centering
        \includegraphics[width=\textwidth]{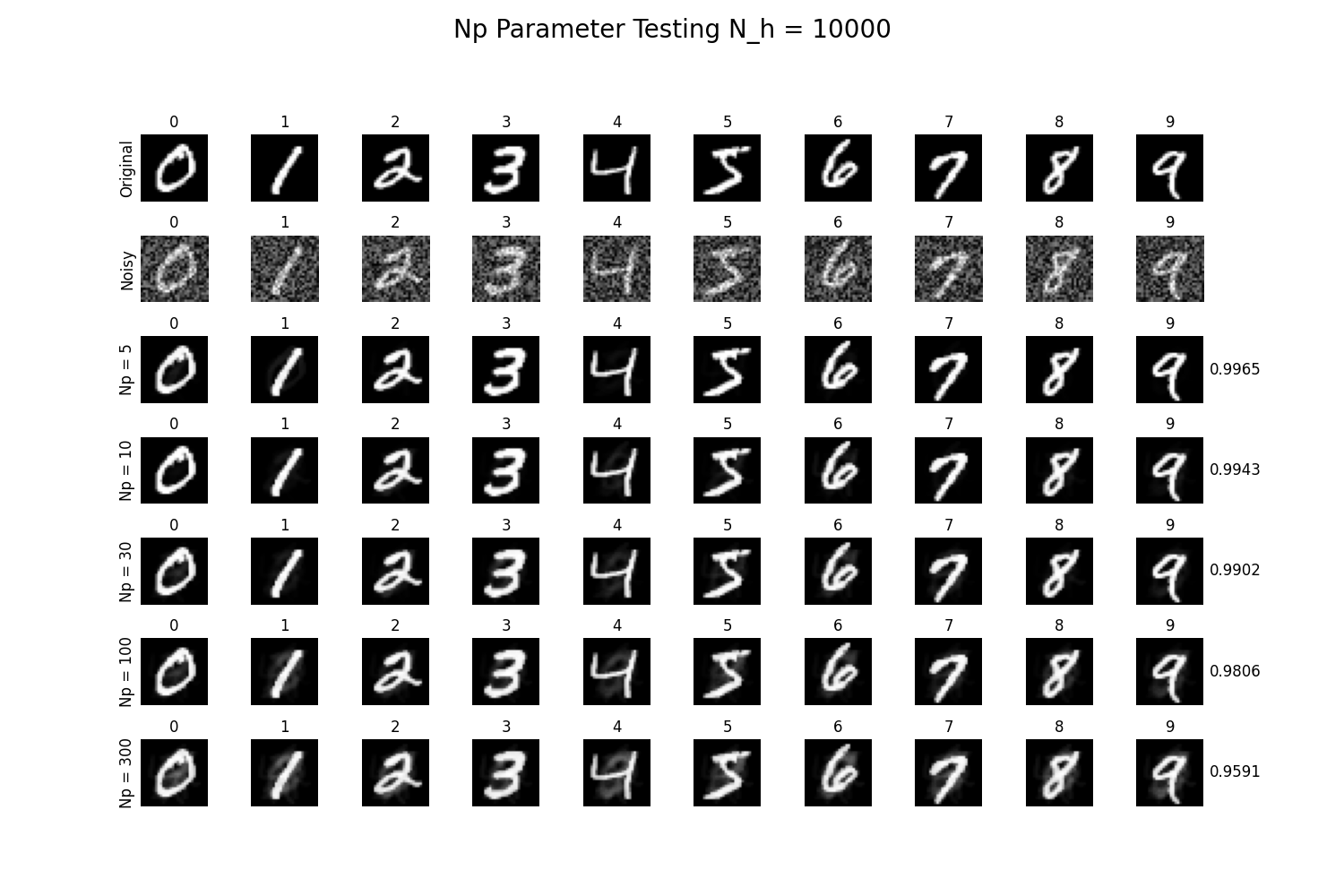}
        \caption{Recall Performance with $N_h$ = 10000.}
        \label{p_test_10000}
    \end{subfigure}
    \hfill
    \begin{subfigure}{0.4\textwidth}
        \centering
        \includegraphics[width=\textwidth]{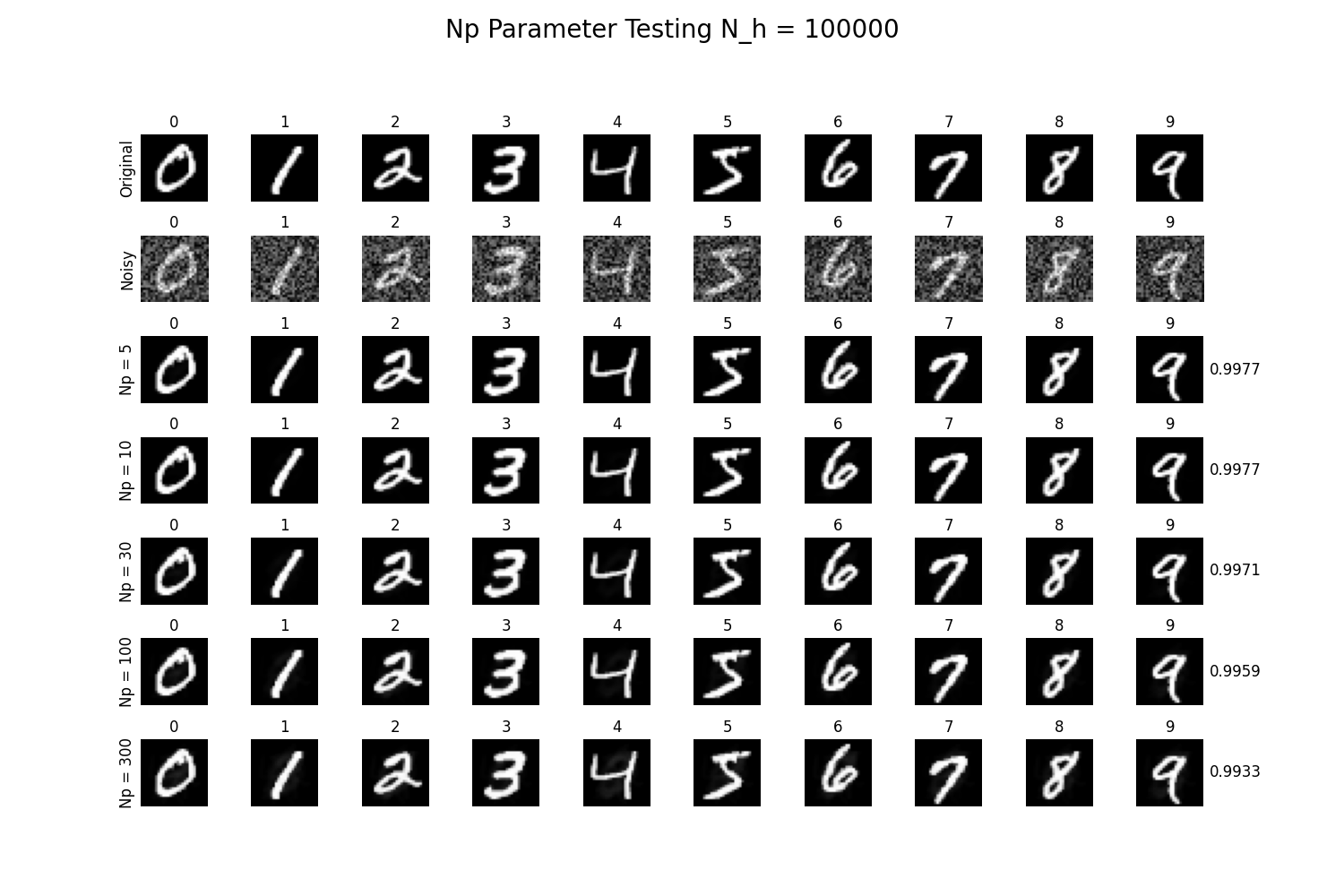}
        \caption{Recall Performance with $N_h$ = 100000.}
        \label{p_test_100000}
    \end{subfigure}
    
    \caption{NP parameter test with hyperparameters: lambdas = [11, 13, 17], path = straight line, Path Length: 30, Number of Images = 10, Corruption = 'medium', grid type = Square.}
    \label{fig:p_parameter_test}
\end{figure}

\begin{figure}
    \centering \includegraphics[width=\linewidth]{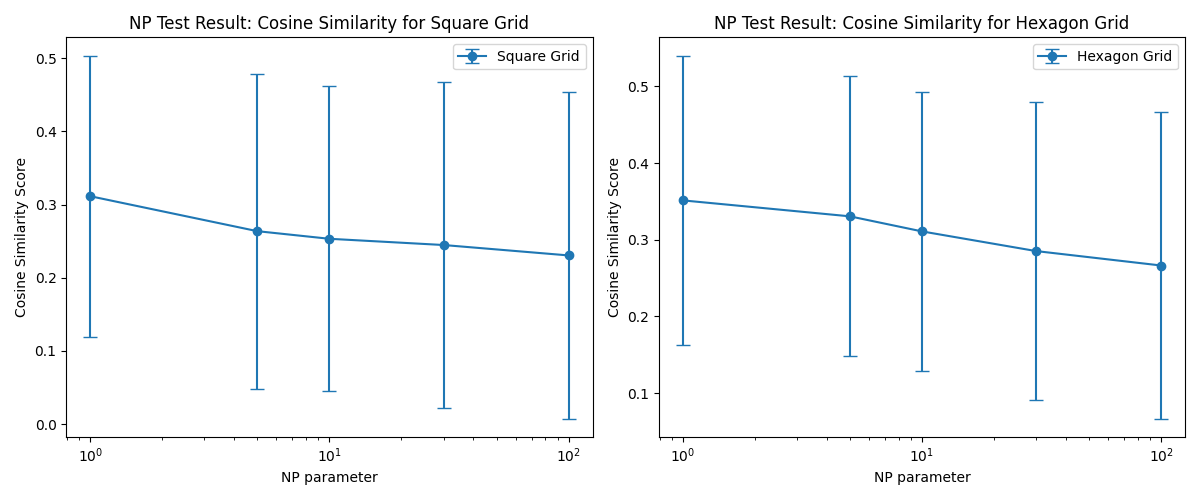}
    \caption{Np test for values: Np = 1, 5, 10, 30, and 100. Hyperparameters: $lambda = (1,2,3)$, dataset = CIFAR100, path = lévy, $N_h = 100000$, $\text{number of images = 455}$}
    \label{fig:nptest}
\end{figure}

\begin{figure}
    \centering \includegraphics[width=\linewidth]{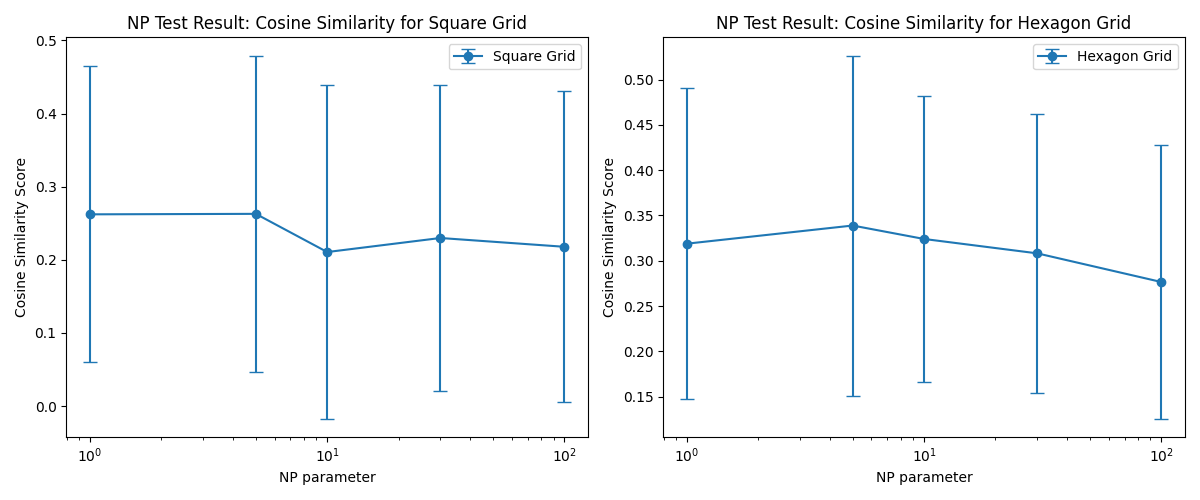}
    \caption{Second Np test for values: Np = 1, 5, 10, 30, and 100. Hyperparameters: $lambda = (1,2,3)$, dataset = CIFAR100, path = lévy, $N_h = 100000$, $\text{number of images = 455}$}
    \label{fig:nptestmain2}
\end{figure}

\clearpage
\end{document}